\documentclass[a4paper,10pt,onecolumn]{article}
\usepackage{graphicx}
\usepackage{natbib}
\usepackage{txfonts}
\usepackage{hyperref}
\usepackage{rotating}
\usepackage{adjustbox}
\usepackage{listings}
\usepackage{xcolor}

\title{ESO/HARPS Radial Velocities Catalog}

\author{Mauro Barbieri$^{1,2}$\\
{\small ORCID: \href{https://orcid.org/0000-0001-8362-3462}{0000-0001-8362-3462}}\\
~\\
{\scriptsize 1: Terma GmbH, Europaarkaden II, Bratustrasse 7, 64293 Darmstadt, Germany}\\
{\scriptsize 2: ESO - European Southern Observatory, Karl-Schwarzchild-Strasse 2, 85748 Garching bei M\"unchen, Germany}\\
~\\
}
\date{11 December 2023}

\begin{document}
\maketitle
\clearpage

\section{Abstract}

This document details the first public data release of the HARPS radial velocities catalog.
This data release aims to provide the astronomical community with a catalog of radial velocities obtained with spectroscopic observations acquired from 2003 to 2023 with the High Accuracy Radial Velocity Planet Searcher (HARPS) spectrograph installed at the ESO 3.6m telescope in La Silla Observatory (Chile), and spanning wavelengths from 3\,800 to 6\,900 \AA.
The catalog comprises 289\,843 observations of 6\,488 unique astronomical objects.

Radial velocities reported in this catalog are obtained using the HARPS pipeline, with a typical precision of 0.5 m/s, which is essential for the search and validation of exoplanets. Additionally, independent radial velocities measured on the H$\alpha$ spectral line are included, with a typical error of around 300 m/s suitable for various astrophysical applications where high precision is not critical.
This catalog includes 282\,294 radial velocities obtained through the HARPS pipeline and 288\,972 derived from the H$\alpha$ line, collectively forming a time-series dataset that provides a historical record of measurements for each object.

Further, each object has been cross-referenced with the SIMBAD astronomical database to ensure accurate identification, enabling users to locate and verify objects with existing records in astronomical literature. Information provided for each object includes: astrometric parameters (coordinates, parallaxes, proper motions, radial velocities), photometric parameters (apparent magnitudes in the visible and near-infrared), spectral types and object classifications.

\section{Introduction}

Over the past 20 years, from June 2003 to June 2023, more than 300\,000 observations have been made at ESO\footnote{\url{https://www.eso.org/}} La Silla Observatory\footnote{\url{https://www.eso.org/sci/facilities/lasilla.html}}, using the 3.6m telescope\footnote{\url{https://www.eso.org/sci/facilities/lasilla/telescopes/3p6.html}} equipped with the HARPS instrument\footnote{\url{https://www.eso.org/sci/facilities/lasilla/instruments/harps.html}}, and have greatly contributed to the fields of extrasolar planets research and stellar physics.

Despite the high precision of the observational data, the recompilation of a catalog of the observations face some significant issues relating the identification of the observed targets due to metadata inconsistency concerning the names of the observed targets stored in the FITS files and to the values of equatorial coordinates stored in the same files.

To address these concerns and in order to provide the users a catalog observations with accurate coordinates and consistent target names,
a comprehensive curation effort has been undertaken to clean and standardize the metadata associated with the HARPS observations. A brief review of the issues found in the HARPS files metadata is provided in Appendix \ref{app_metadata}.

\section{The 3.6-m Telescope and HARPS}

\subsection{The 3.6-m Telescope}

The ESO 3.6-m telescope, is an equatorial mount telescope with a diameter of 3.566 meters and a focal ratio of F/8, that was built in the '70s in collaboration between ESO and CERN, and saw its first light in 1976. It is located at the La Silla Observatory in Chile, and the telescope coordinates (in WGS 84 system\footnote{
The WGS 84 standard, maintained by the United States National Geospatial-Intelligence Agency~(\url{https://earth-info.nga.mil}), establishes an Earth-centered, Earth-fixed coordinate system. It provides a global reference frame for accurately representing locations on Earth through geodetic coordinates, which are a specific type of curvilinear orthogonal coordinate system based on a reference ellipsoid. Additionally, WGS 84 details the Earth Gravitational Model (EGM) and World Magnetic Model (WMM).
}) are:\\
\begin{table}[h]
\begin{tabular}{llll}
Latitude &:&$-29^\circ 15' 39.5"$ S& (-29.260972$^\circ$ S)\\
Longitude&:&$-70^\circ 43' 54.1"$ W& (-70.731694$^\circ$ W)\\
Elevation&:&2400 m above sea level&\\
\end{tabular}
\end{table}
\noindent
~\\
It is important to note that the telescope coordinates reported here do not correspond to the coordinates in the FITS headers of any instrument installed on this telescope. The origin of the erroneous coordinates in the FITS header is unclear and is currently under investigation. Despite the potential impact of coordinate differences on the calculation of the Barycentric Earth Radial Velocity (BERV) correction, we have verified that the radial velocity signal introduced by this discrepancy is on the order of 2 cm/s peak to peak with a period of one sidereal day.

~\\
The pointing procedure of the telescope can acquire a target with a typical error of 5 arcsec, this accuracy slightly diminishes when pointing to the North. The autoguiding system maintains an accuracy of less than 0.1 arcsec. The image quality of the telescope is better than 0.2 arcsec at Zenith.
Since HARPS became operational in 2003, the telescope has undergone the following significant interventions:
\begin{itemize}
\item
in 2004, a new secondary mirror cell was manufactured, which significantly improved the image quality of the telescope
\item
the main mirror was coated with aluminium in the years 2005, 2008, 2014, 2017, and 2021.
\end{itemize}

\noindent
More technical details of the telescope and of his history can be found here:
\begin{itemize}
\setlength\itemsep{-0.2em}
\item in the telescope science page:\\
\url{https://www.eso.org/sci/facilities/lasilla/telescopes/3p6.html}
\item in the telescope public page:\\
\url{https://www.eso.org/public/teles-instr/lasilla/36}
\item actual and historical instruments information, are available from the telescope public page in the section "Telescopes \& Instruments /\ La Silla Observatory/\ ESO 3.6-metre telescope" accessible from the main menu
\item in the ESO annual reports:\\
\url{https://www.eso.org/public/products/annualreports}
\item in the ESO bulletins:\\
\url{https://www.eso.org/public/products/bulletins}
\item in the book {\it ESO's Early History} by Adriaan Blaauw (chapters 8 and  9)\\
\url{https://www.eso.org/sci/libraries/historicaldocuments/ESO_Early_History_Blaauw/ESO_Early_History.pdf}
\end{itemize}

\subsection{HARPS}

HARPS, is a fiber-fed echelle spectrograph optimized for extreme radial velocity accuracy, targeting very high long term radial velocity accuracy (on the order of 1 m/s) installed at ESO 3.6-m telescope on the F/8 Cassegrain focus. It achieves this high accuracy thanks to the installation of all the optical system in a vacuum vessel that mitigates temperature and pressure-induced spectral drifts and a dual-fiber system (1 arcsec aperture) for target and either Th-Ar reference spectrum (or sky background). The system covers a spectral range of 3780-6910 \AA, with a resolution of R=115\,000. The detector consists of a 4k x 4k mosaic of 2 CCDs, losing one spectral order due to the gap between chips (from 5300 to 5330 \AA).

HARPS attains a signal-to-noise ratio of 110 per pixel at 5500 \AA\,for a G2V star of V magnitude = 6, in 1 minute integration time (1 arcsec seeing, airmass = 1.2).
With this signal-to-noise ratio and using the Simultaneous Thorium Reference Method, and considering errors introduced by guiding, focus, and instrumental uncertainties, the overall radial velocity accuracy is about 1 m/s.

HARPS is equipped with its pipeline, that provides extracted and wavelength calibrated spectra for all the HARPS observing modes.

\begin{table}[h!]
\begin{center}
\begin{tabular}{|l|l|}
\hline
\multicolumn{2}{|c|}{HARPS main characteristics}\\
\hline
Spectral range & 3780-6910 \AA\\
Spectral format & "red" CCD (Jasmin): orders 89-114, 5330-6910 \AA \\
&"blue" CCD (Linda): orders 116-161, 3780-5300 \AA\\
Spectral resolution & 115\,000 \\
Sampling per spectral element & 3.4 px per FWHM\\
Spectrum Separation & 17.3 px (fibers A and B)\\
Order separation "red" & order 89: 100.7px, order 114: 62.7px\\
Order separation "blue" & order 116: 60.7px, order 161: 34.2px\\
\hline
\end{tabular}
\end{center}
\label{tab01a}
\end{table}

\begin{center}
\begin{figure}
    \includegraphics[width=1.05\columnwidth]{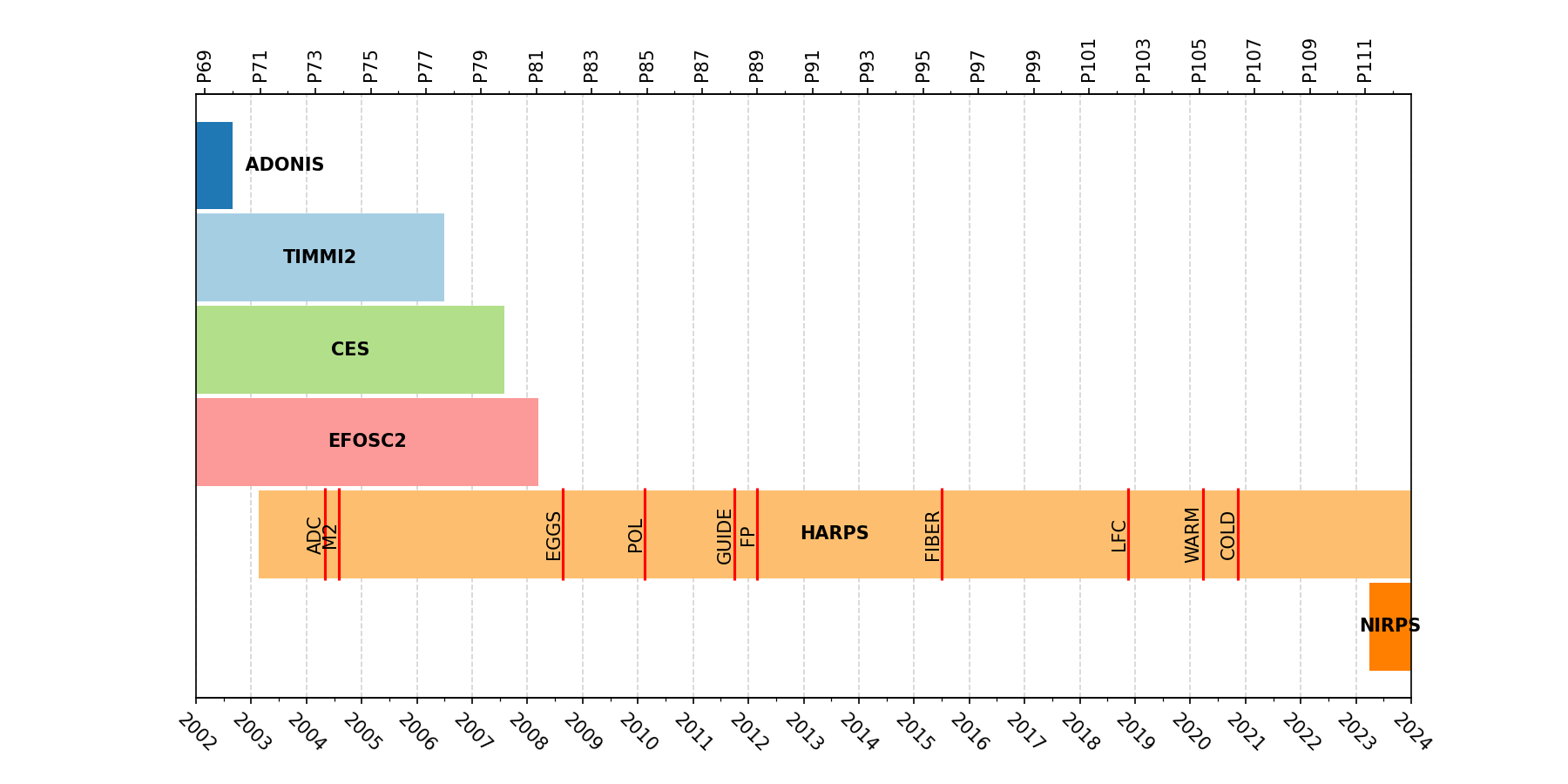}
    \caption{Timeline of instrumentation availability from first light of HARPS till the end of 2023.
    In the upper horizontal axis shows the tics represents the beginning of an ESO observing period.}
    \label{f:timeline}
\end{figure}
\end{center}

From the installation of HARPS in 2003 until 2008, it has shared the telescope with only other 3 instruments (CES, TIMMI2 and EFOSC2),
from 2008 to 2023 it was the only instrument on the telescope,
and from April 2023 it is sharing the telescope with NIRPS.
In Fig\ref{f:timeline} is shown the graphical timeline
of the instrument at the telescope from the installation of HARPS.

HARPS underwent several significant updates and changes since its first light on February 11, 2003; key changes include:
\begin{itemize}
\item
the installation of an Atmospheric Dispersion Corrector (ADC) in May 2004
\footnote{\url{https://www.ls.eso.org/sci/facilities/lasilla/instruments/ces/headers.html}}
\item
the Iodine self-calibration method was taken out of service starting from May 2004
\footnote{\url{https://www.ls.eso.org/lasilla/Telescopes/3p6/harps/manuals/userman1_3.pdf}}
\item
the high efficiency fiber link (EGGS) was added in 2006.
\item
in August 2008 the reference fiber for EGGS was discovered damaged and not transmitting light\footnote{\url{https://www.eso.org/sci/facilities/lasilla/instruments/harps/eggs-aug2008.html}}, it was replaced in 2015 (see below).
\item
the spectro-polarimeter capability was added in 2010 \citep{2011Msngr.143....7P}, the corresponding science data products can be downloaded from  here:\\
\url{http://archive.eso.org/wdb/wdb/eso/repro/form}.
\item
in 2015 the original circular shaped optical fibers were replaced by new octagonal fibers \citep{2015Msngr.162....9L}.
This upgrade led to the elimination of systematic due to de-centering and de-focusing, repaired the broken sky fiber of EGGS, and increased the throughput of HARPS by at least 10\%, improving light scrambling capabilities and increasing the stability of the light injection within the spectrograph.
The installation of the new fibers, leaded to a shift in radial velocity before and after the date of the installation. A brief resume of the fiber related problematics is provided in Appendix \ref{app_fiber}.

\item
the Laser Frequency Comb (LFC) started running at HARPS in 2018, but it is used only for calibrations and not offered for night-time observations,
for more details see \cite{2012Msngr.149....2L}.
HARPS data obtained during commissioning run of LFC can be downloaded from here:\\
\url{https://www.eso.org/sci/activities/instcomm/harps_lfc.html}

\item
On 21 Oct 2014, the processed HARPS echelle data were made available as Science Data Products (SDP)\footnote{\url{https://www.eso.org/sci/observing/phase3.html}} through the Phase 3 query form%
\footnote{\url{http://archive.eso.org/wdb/wdb/adp/phase3_main/form?phase3_collection=HARPS}}
and spectral query form%
\footnote{\url{http://archive.eso.org/wdb/wdb/adp/phase3_spectral/form?phase3_collection=HARPS}}
, and trough the ESO Science portal%
\footnote{\url{https://archive.eso.org/scienceportal/home?data_collection=HARPS}}.
This dataset contain HARPS DRS pipeline products: 1D spectra in the ESO/SDP PHASE3 standard format and the ancillary tar files.
The dataset spans from HARPS inception in October 2003 to the present.
\end{itemize}

\noindent
Additional information on the HARPS instrument can be found in the following documents:
\begin{itemize}
 \item HARPS User Manual \cite{3P6-MAN-ESO-90100-0005}: this document describe all necessary information of the users of the HARPS instrument
 \item HARPS DRS User Manual \cite{3M6-MAN-HAR-33110-0016}: this document describes the Data Reduction Software (DRS) its architecture and configuration
 \item HARPS Templates Reference Document \cite{3M6-TRE-HAR-33110-0008}: this document describes the templates needed for the scientific, calibration and maintenance operations of HARPS
\end{itemize}

\section{Overview of Catalog}

In this data release are included:
\begin{itemize}
\item Data regarding radial velocities determination with the Cross-Correlation Functions (CCF) and lines Bisectors, as computed by the HARPS Data Reduction Software (DRS), were retrieved from the ancillary ADP files and are included in this dataset.
\item Independent radial velocities obtained from the H$\alpha$ line at 6562.79\AA\ for targets lacking HARPS DRS measurements. Given the nature of the H$\alpha$ line, these velocities have a typical error margin of 300 m/s, making them less suitable for exoplanet detection but valuable for other astrophysical applications.
\item A correlation of HARPS observation coordinates with the SIMBAD astronomical database \citep{2000A&AS..143....9W} for consistent identification of observed targets (\url{http://simbad.cds.unistra.fr/simbad}).
\end{itemize}

\begin{figure}
 \centering
 \includegraphics[width=1.5\columnwidth,angle=-90,keepaspectratio=true]{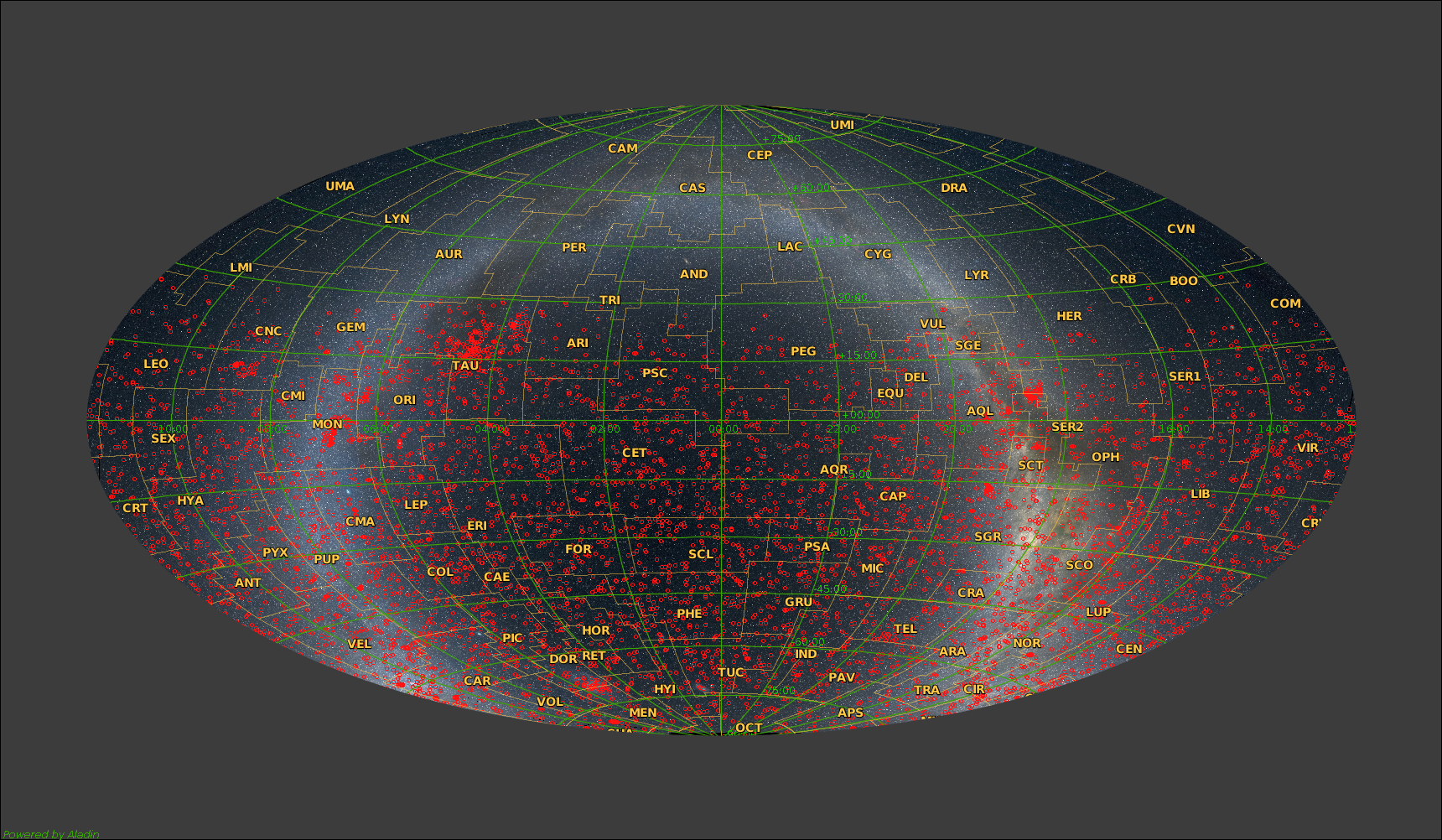}
 \caption{Sky coverage of HARPS observations in equatorial coordinates. Each circle represent a single observation. The image was created using Aladin sky atlas \citep{2000A&AS..143...33B} and the all sky background image is taken from \cite{2009PASP..121.1180M}.}
 \label{fig:01aa}
\end{figure}

\begin{figure}
 \centering
 \includegraphics[width=1.5\columnwidth,angle=-90,keepaspectratio=true]{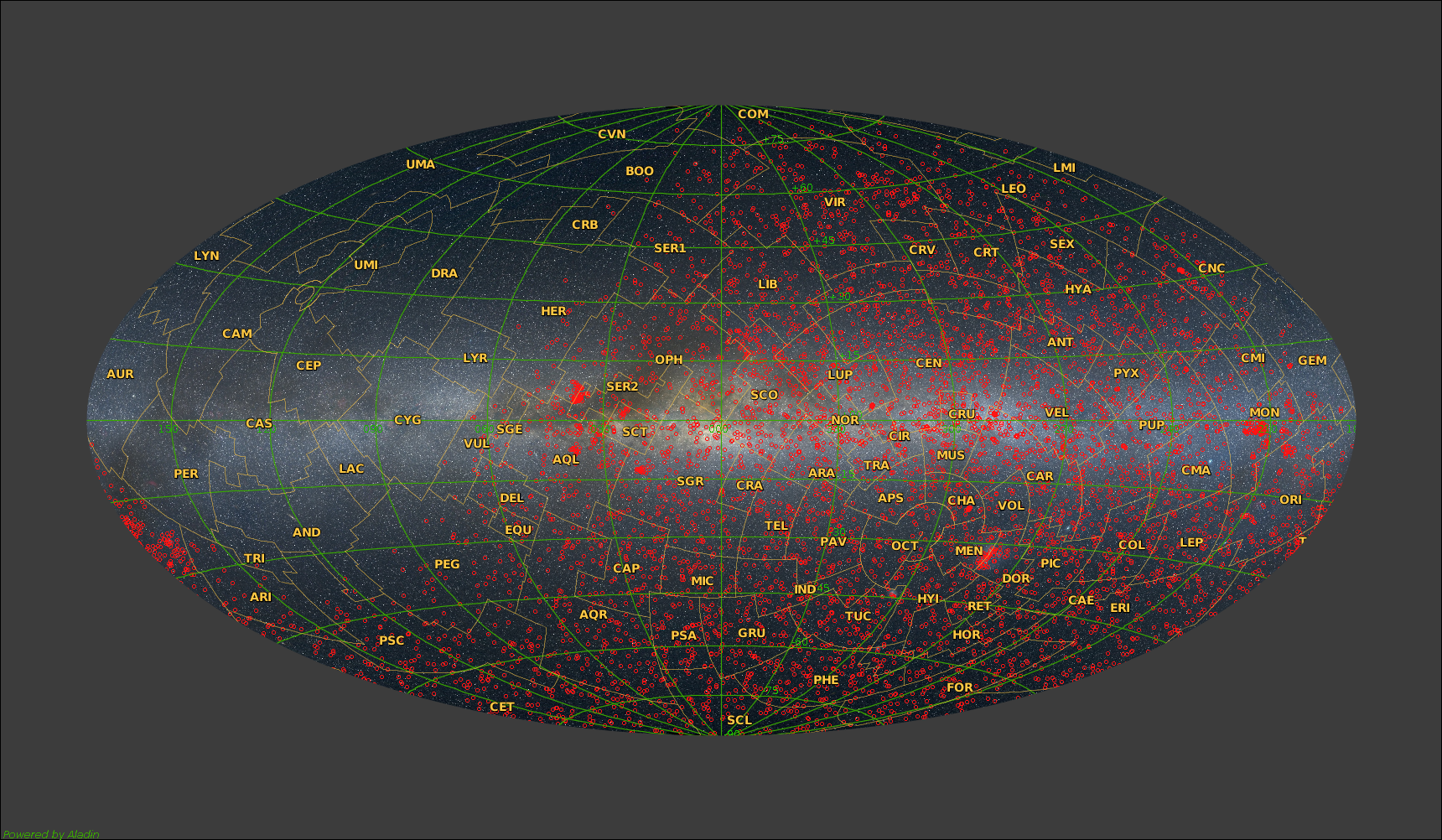}
 \caption{Sky coverage of HARPS observations in galactic coordinates. Each circle represent a single observation. The image was created using Aladin sky atlas \citep{2000A&AS..143...33B} and the all sky background image is taken from \cite{2009PASP..121.1180M}.}
 \label{fig:01ab}
\end{figure}

In total, HARPS has been utilized in 630 programs by 327 different principal investigators and co-investigators. The dataset contains 289\,843 observations, with their celestial distribution depicted in Fig.\ref{fig:01aa} and Fig.\ref{fig:01ab}. Areas of concentrated observations correspond to exoplanet transit follow-up fields or the Magellanic clouds.

\begin{figure}
 \centering
 \includegraphics[height=0.27\textheight]{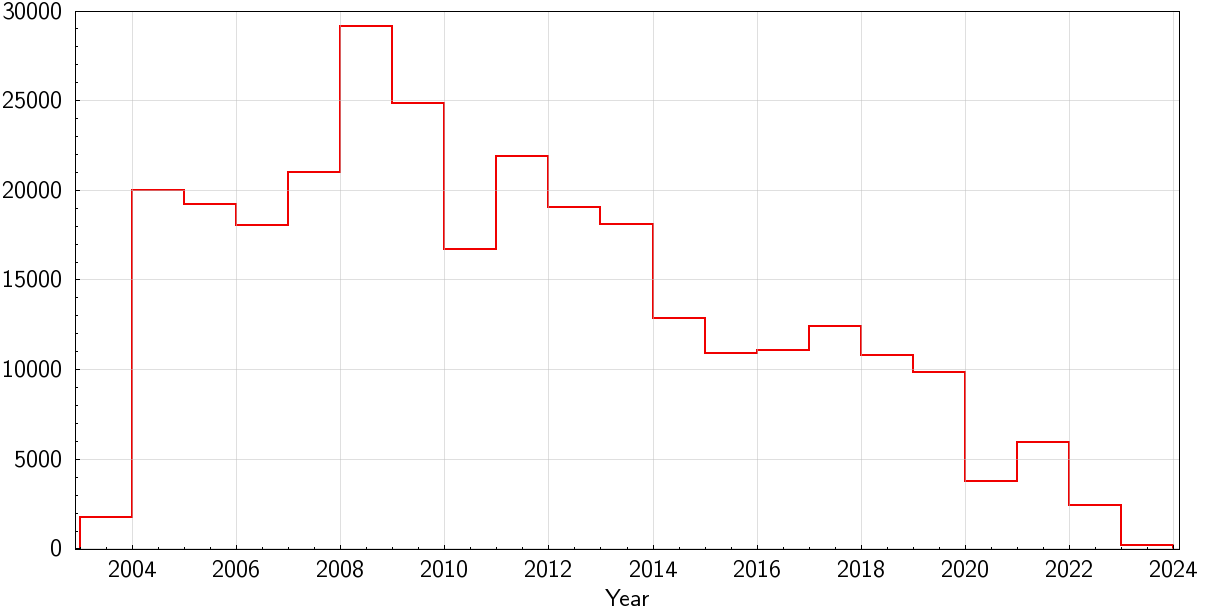}
 \caption{Number of HARPS observations per year.}
 \label{fig:02}
 \includegraphics[height=0.27\textheight]{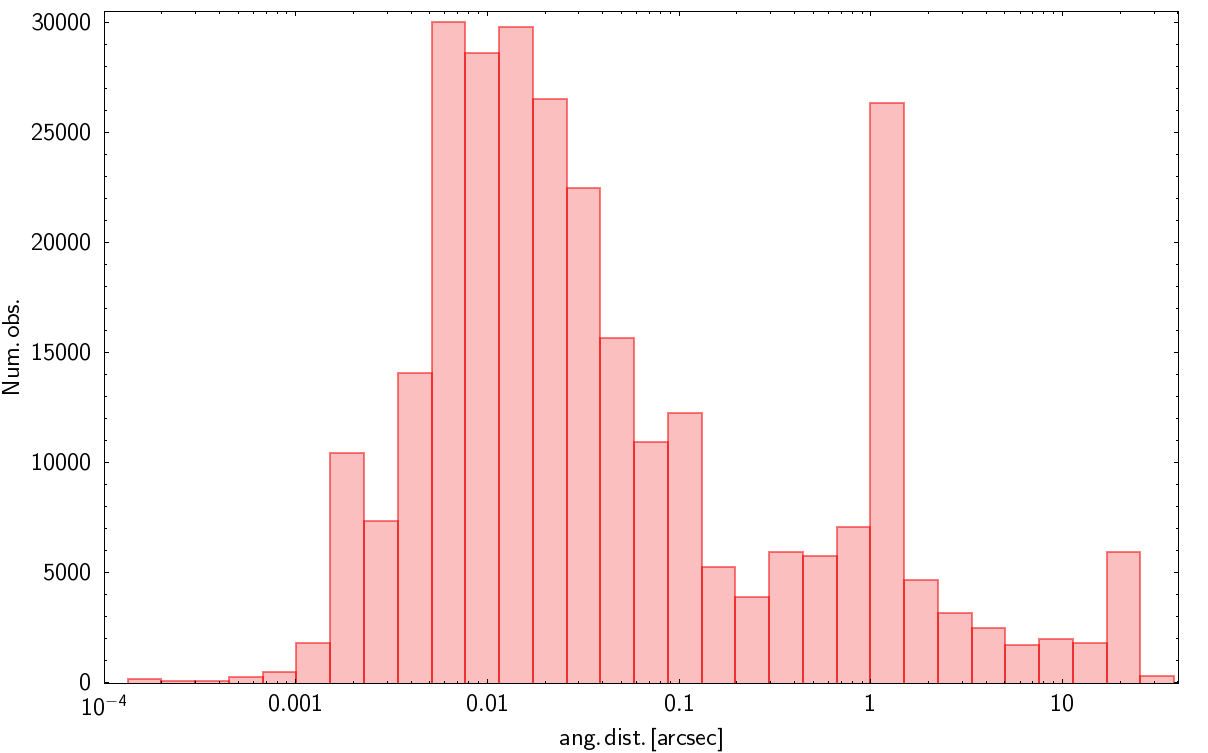}\\
 \includegraphics[height=0.27\textheight]{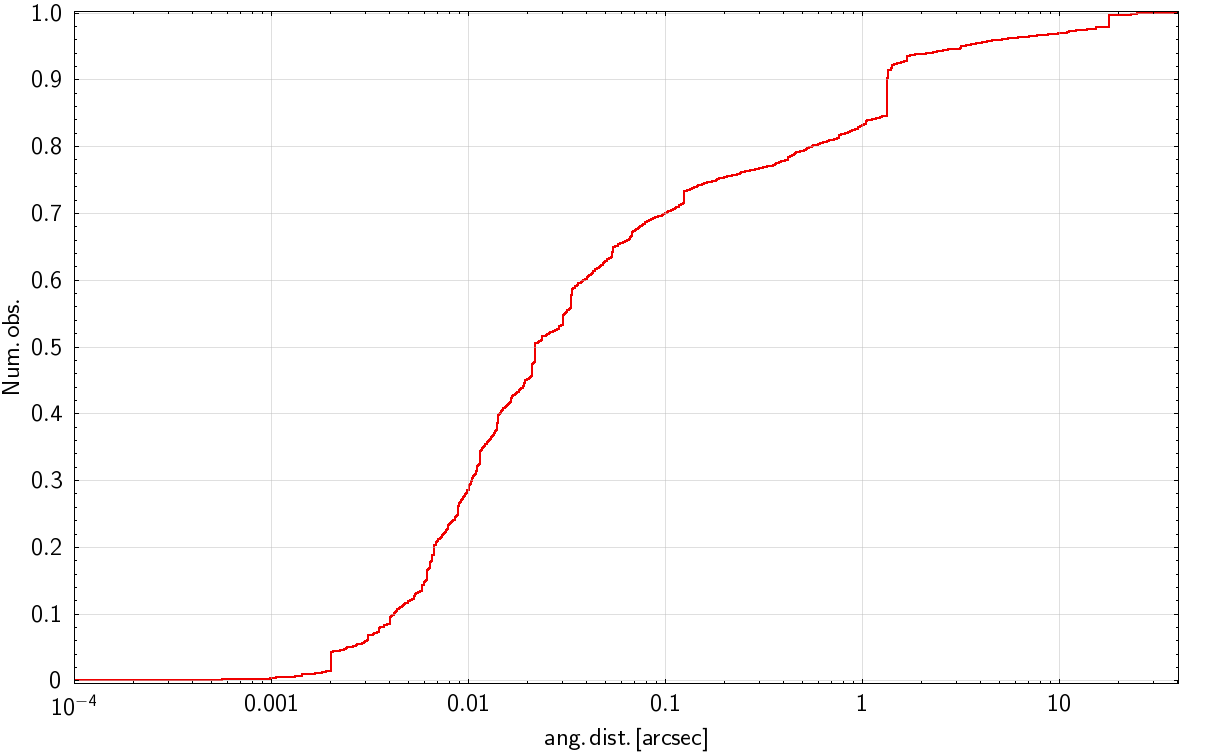}
 \caption{Angular distance between the observed targets and the SIMBAD entry. Top panel: histogram of the distribution. Bottom panel: cumulative distribution.}
 \label{fig:03}
\end{figure}

The mean annual number of observations is illustrated in Fig.\ref{fig:02}, shows a decrease over time, reflecting the trend towards observing fainter targets requiring longer exposure times and operational impacts during the COVID-19 pandemic.

A successful match with SIMBAD was achieved for 286\,767 observations (99\% of the dataset), with a detailed angular distance distribution between observed targets and SIMBAD entries presented in Fig.\ref{fig:03}. Notably, over 91\% of the matches occurred within 2 arcsec. Only 3\,077 observations do not have a counterpart in SIMBAD within 30 arcsec of the target coordinates.

\begin{figure}
 \centering
 \includegraphics[width=\columnwidth,keepaspectratio=true]{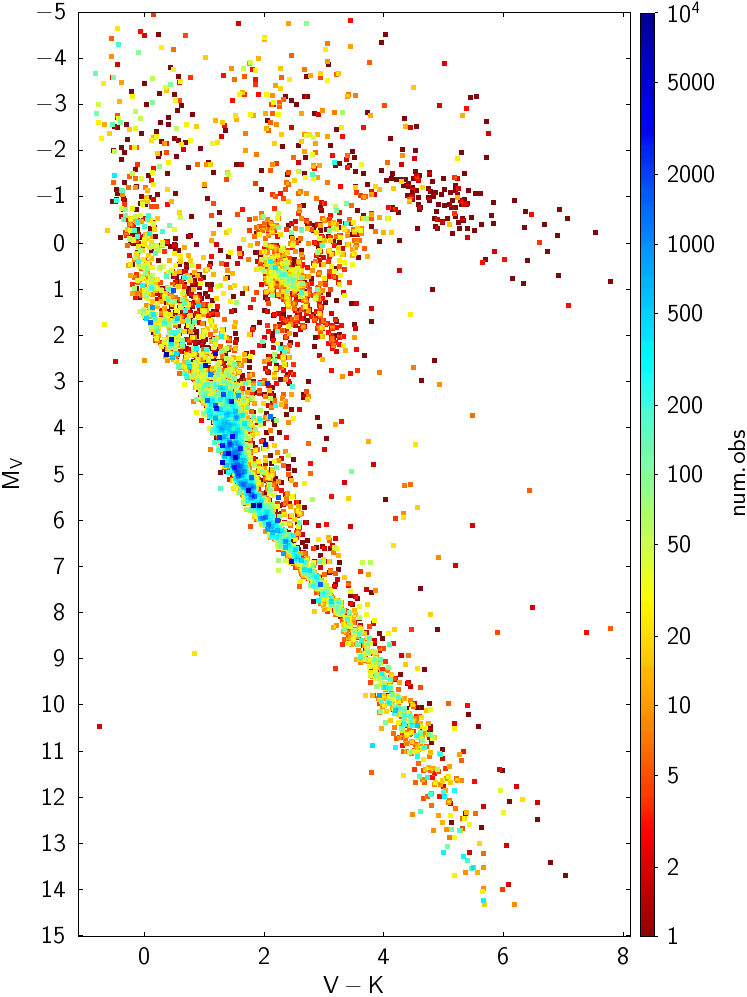}
 \caption{HR diagram of the observed targets. The abscissa provide the V-K color from SIMBAD (without extinction correction), while the ordinate provide the absolute magnitude calculated from SIMBAD: M$_{\rm{V}}$ = Vmag - 5log(1/plx)+5.  The color scales indicate the number of observations per target.}
 \label{fig:04}
\end{figure}

The targets observed by HARPS span across almost the entire Hertzsprung-Russell (HR) diagram (see Fig.\ref{fig:04}) and most of the observed targets are concentrated in the main sequence between absolute magnitudes M$_{\rm V} =$ 4 and 6.

\begin{figure}
 \centering
 \includegraphics[width=\columnwidth]{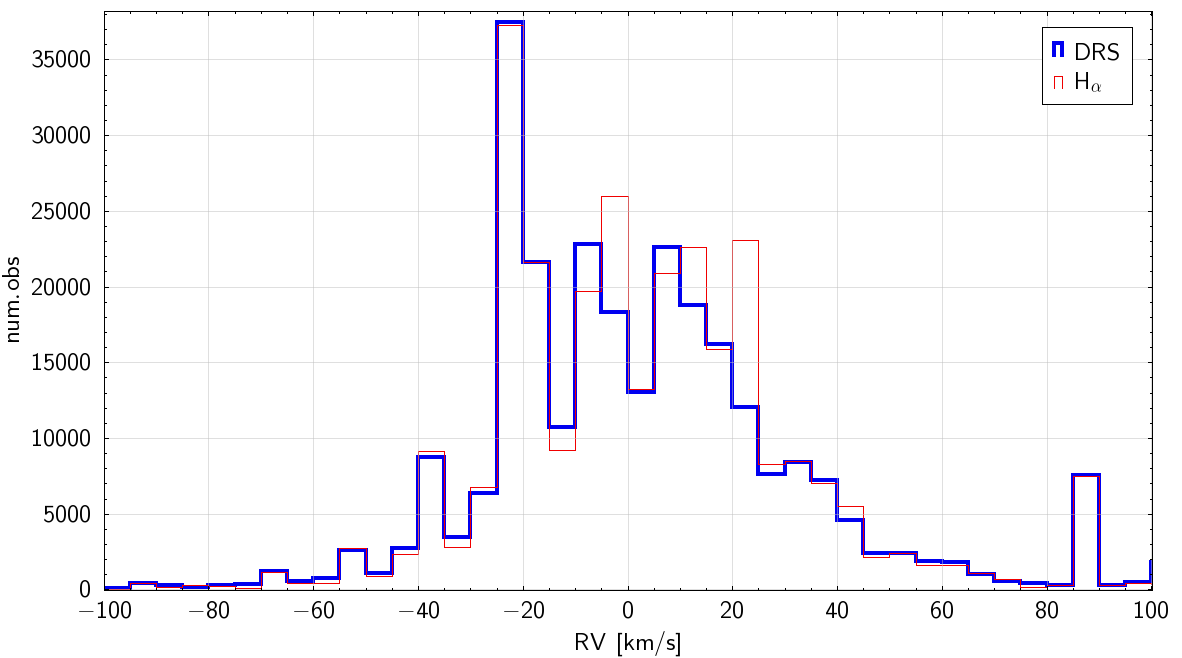}
 \caption{Histogram of the radial velocities measurements from the HARPS DRS (in blue) and the H$\alpha$ line (in red).}
 \label{fig:05}
\end{figure}

\begin{figure}
 \centering
 \includegraphics[height=0.27\textheight]{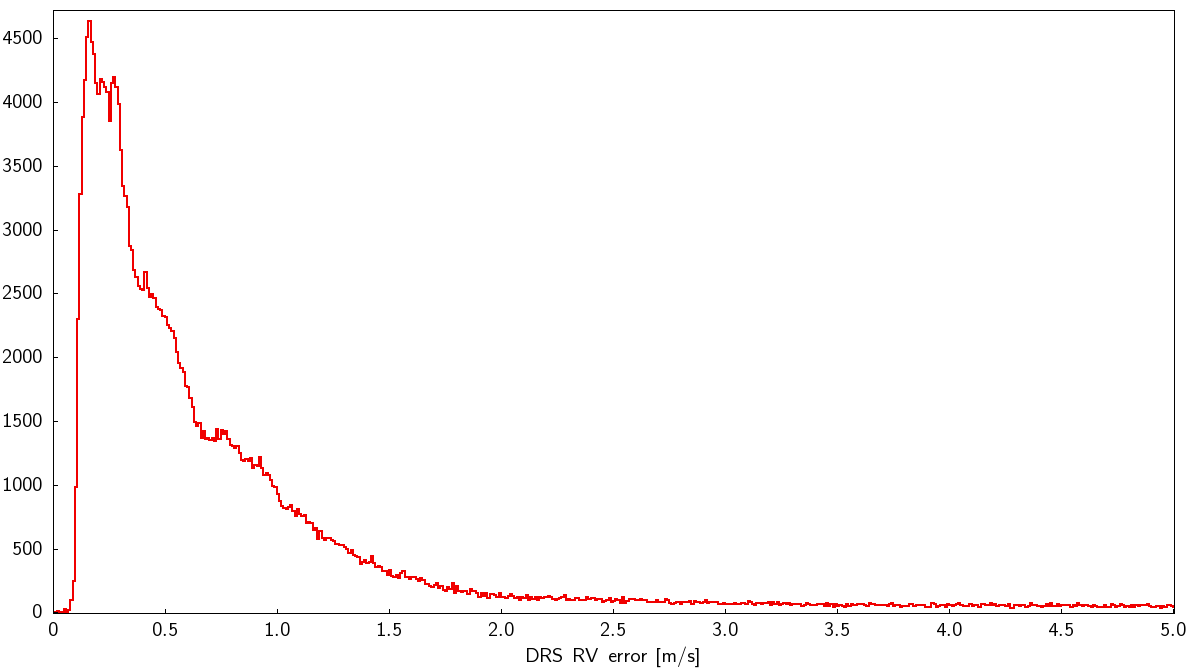}\\
 \includegraphics[height=0.27\textheight]{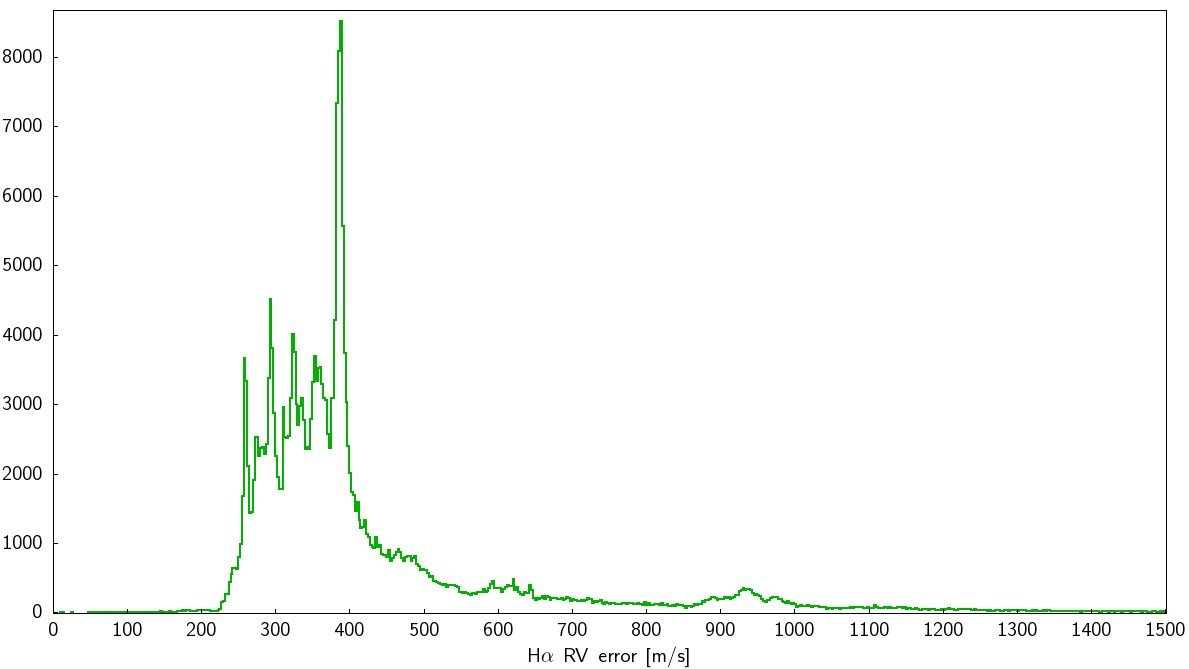}
 \caption{Histograms of the radial velocities errors from the HARPS DRS and H$\alpha$ line.}
 \label{fig:05a}
\end{figure}

Fig.\ref{fig:05} shows the distribution of radial velocities measured by the HARPS DRS and from the H$\alpha$ line. Fig.\ref{fig:05a} presents the corresponding distribution of measurement errors for both methods.

Out of the total dataset, 11\,900 observations (4\%) have HARPS DRS-measured radial velocities that are either above 1000 km/s in absolute value or are missing. In contrast, only 677 observations lack H$\alpha$ radial velocity measurements or display values exceeding 1\,000 km/s in absolute value.

\begin{figure}
 \centering
 \includegraphics[width=0.5\columnwidth]{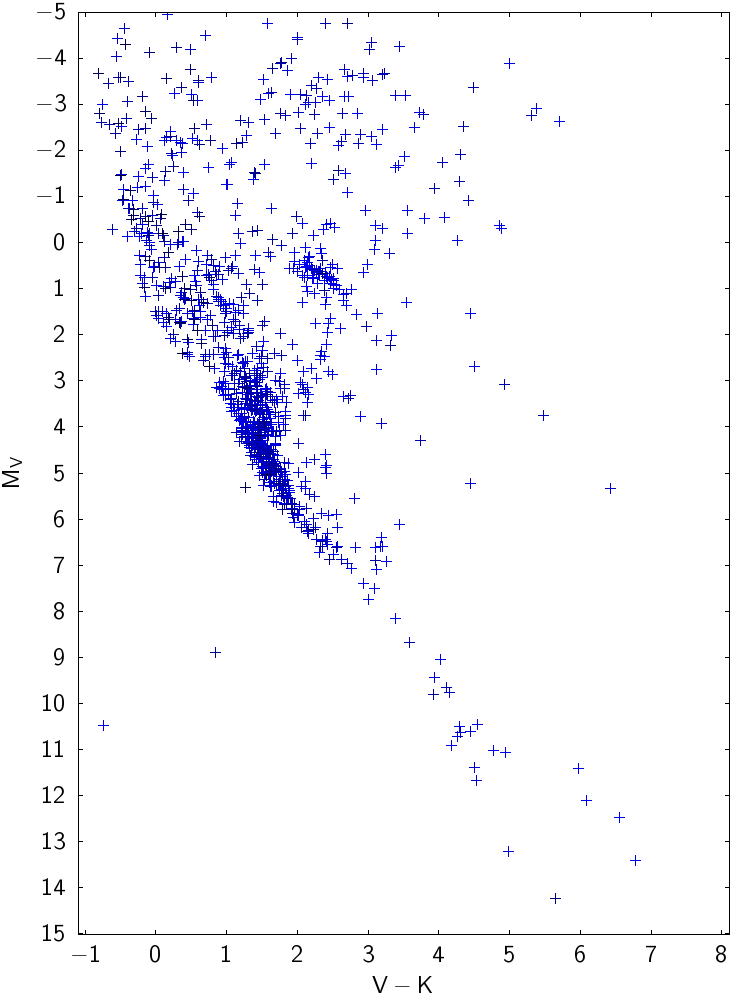}%
 \includegraphics[width=0.5\columnwidth]{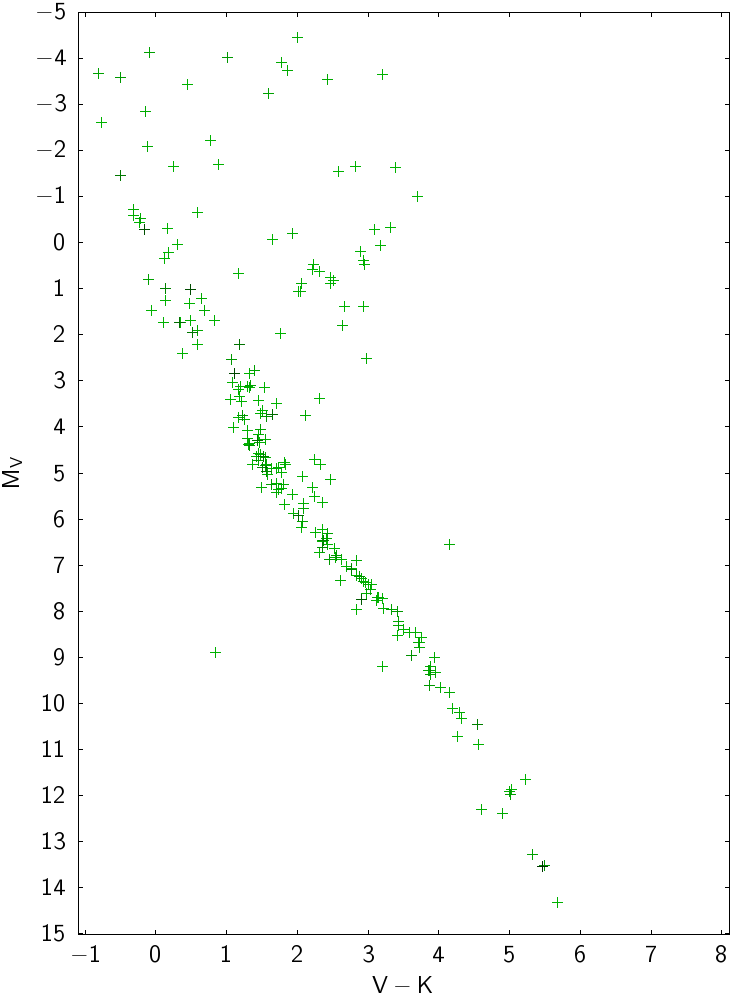}
 \caption{Left panel: HR diagram of the targets without RV measurements from the HARPS DRS.
 Right panel: HR diagram of the targets without RV measurements from H$\alpha$ line.\\}
 \label{fig:06}
 \includegraphics[width=\columnwidth]{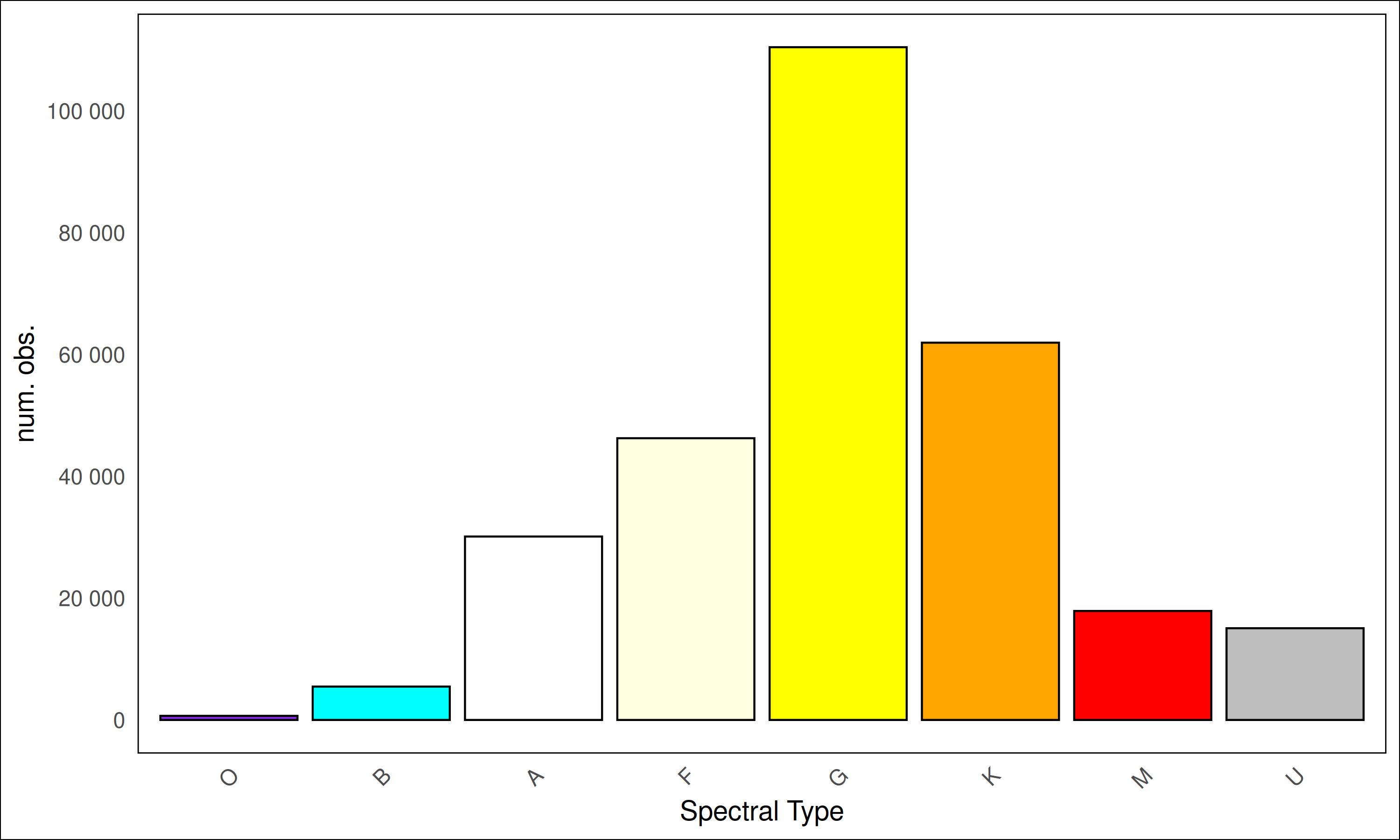}
 \caption{Histogram of spectral types from SIMBAD classification. The gray bar correspond with target with non-standard classification or missing classification.}
 \label{fig:06b}
\end{figure}

Within the total dataset, 11\,900 observations (4\%) have HARPS DRS-measured  radial velocities either exceeding 1\,000 km/s or are missing. Conversely, just 677 observations are without H$\alpha$ radial velocity measurements or have values above 1\,000 km/s. The 1\,000 km/s is a generous margin for the maximum galactic radial velocities that could be observed by HARPS.
Figure \ref{fig:06} illustrates the HR diagrams for these subsets, underscoring the distribution of absent radial velocities. Notably, HARPS DRS-derived radial velocities are predominantly missing for stars earlier than type F, whereas H$\alpha$-derived velocities, despite their lower precision, are available across all spectral types.

The broader availability of H$\alpha$ radial velocities makes them a versatile tool for a wider range of astrophysical investigations beyond the scope of exoplanet detection.

Finally Figure \ref{fig:06b} show the number of observations per spectral types where is clear that the core of the observation performed on HARPS is on G-type stars.

\section{Catalog Generation}

This data release covers the first 20 years of observations by the High Accuracy Radial Velocity Planet Searcher (HARPS) \citep{2003Msngr.114...20M}, spanning from its initial operation in June 2003 to June 2023.

This release excludes observations that utilized the polarimetric setup, lasercomb setup, and spectra pertaining to Solar System objects and the Sun. These data will be incorporated in subsequent releases.

We retrieved the data for building this catalog from the ESO science archive, specifically from the HARPS Phase3 data collection\footnote{
Data: \url{https://archive.eso.org/scienceportal/home?data_collection=HARPS} \
Documentation: \url{https://www.eso.org/rm/api/v1/public/releaseDescriptions/72}\\
DOI: \url{https://doi.eso.org/10.18727/archive/33}\\
}

The Radial Velocities of HARPS are stored in FITS files within the ancillary file  of the science file (ADP). These ancillary files are TAR archives containing intermediary products of the HARPS pipeline. The format of the science ADP files and the content of the ancillary files are described in \cite{harps..stream..2023}.

Retrieving ADP files from the ESO Archive is straightforward, unlike the ancillary files. To address this, we developed a procedure using TAP queries on the ESO TAP service to associate ADP files with ancillary and RAW files. This process is detailed in a Python script available in Appendix \ref{app_code}.

We downloaded all available HARPS ADP files up to June 2023 to mark the 20th anniversary of its first public dataset. For each ADP file, we downloaded and decompressed the corresponding ancillary TAR file. We then extracted relevant keywords from the ADP, and from the CCF and BIS files, we obtained DRS-related keywords for radial velocities. From the ADP files, we also extracted spectra for measuring the H$\alpha$ radial velocity.

Observations of the Solar System and the Sun were excluded from this compilation. This was done by applying a list of common names to the OBJECT keyword to filter them out. This list is available in Appendix \ref{app_ss}.

\subsection{H$\alpha$ line}

The HARPS spectra due to the instrument static configuration always cover region where is located the H$\alpha$ line at 6562.79\AA\,.
This fact allowed us to use the H$\alpha$ for performing an independent measurement of Radial Velocities (RV).
The measurement was performed fitting a 20\AA\ region centered the theoretical wavelength of the line in air with a Lorentzian absorption  profile:
\begin{equation}
 \rm{Lorentzian}(\lambda) = K - A \cdot \frac{\gamma}{(\lambda - \lambda_0)^{2} + \gamma^{2}}
\end{equation}
Here, K is the value of continuum level, A is the depth of the line, $\gamma$ is the width parameter of the Lorentzian, and $\lambda_0$ is the wavelength of the H$\alpha$ line in air (6562.79\AA), that comes from the NIST tables \citep{2022..NIST}. The quality of the fit is quantified by the reduced $\chi^2$ statistic.

The formula adopted for calculating the H$\alpha$ RV is the non--relativistic Doppler formula:
\begin{equation}
RV = \frac{\lambda_{\rm fit} - \lambda_0}{\lambda_0} c
\end{equation}
where $\lambda_{\rm fit}$ is the value of the center of the line from the fit,
and $c$ is the speed of light ($c =$299\,792.458 km/s, \cite{GGPM..2019..472}).

The associated uncertainty in the radial velocity, is calculated from:
\begin{equation}
\sigma_{RV} = \frac{\lambda_0}{c}\sigma_{\lambda_{\rm fit}}
\end{equation}
where $\sigma_{\lambda_{\rm fit}}$ is the error of the central wavelength as determined from the fit.

\subsection{SIMBAD cross--match}

For the object coordinates, we used the values stored in the keywords HIERARCHICAL\_ESO\_TARG\_ALPHA and HIERARCHICAL\_ESO\_TARG\_DELTA, because these keyword values comes from the observer input when preparing the Observing Blocks.

Using Topcat \citep{2005ASPC..347...29T} we perform a cross--match with SIMBAD using a radius of 30 arcsec and keeping all the matches within this radius. The full list of matches was cleaned to remove objects of the following types: planets, extended sources, clusters, non--optical sources. For performing this cross--match, we utilized the version of SIMBAD available as of June 1, 2023.

A second cleaning was performed in order to remove the binaries stars. More specifically, stars that are multiple in nature are stored in SIMBAD as both individual components (A, B, etc) but also as system of stars. For example $\alpha$ Centauri is stored as the system $\alpha$ Centauri (* alf Cen), and the component A (* alf Cen A) and B (* alf Cen B). We specifically remove all names referring to the systems, but retaining only the names of individual components.

In cases where a match within 30 arcsec of SIMBAD yields a list of objects, despite the previous cleaning step, the final association is made by retaining the closest match from the list, sorted by angular distance.

The detailed process is as follows. The remaining matches were first sorted by the angular distance from the initial cross-correlation. Using Topcat, an internal match was performed utilizing the Exact Value Match Algorithm, which identifies pairs of records across tables sharing an identical key value in a specified column while ignoring pairs with a null key field. The name of the ADP file was used as the key column. We selected the option to 'eliminate all but the first of each group.' Consequently, the algorithm generated a new table by keeping only the first record of each set of duplicates, thereby eliminating subsequent identical entries. An internal match with the same criteria thereafter resulted in no duplicates.

\section{Catalog Content}

In this release, we provide a catalog for HARPS observations between 2003 and 2023, with extracted information from the ADP files and ancillary files, along with H$\alpha$ calculated RV from the spectra and information form SIMBAD on the cross--matched object.
The list of the columns in the tables and their description is provided in the Table \ref{tab01}.
The description of the columns pertaining to the values obtained by the DRS are taken or from the HARPS manuals
(\cite{3P6-MAN-ESO-90100-0005}, \cite{3M6-MAN-HAR-33110-0016}, \cite{3M6-TRE-HAR-33110-0008}) when available, or from the corresponding FITS headers than contain them.

\noindent
The column ins\_det1\_tmmean could be used in conjunction with exptime to find the middle point of the exposure (in seconds):
$$
{\rm mid\_exposure} = {\rm exptime}\cdot {\rm ins\_det1\_tmmean}\quad [s]
$$

\noindent
The list of the most 100 observed objects is provided in Table \ref{t_obs}.

\begin{table}[h]
\centering
\scriptsize
\begin{tabular}{|lr|lr|lr|lr|}
\hline
* alf Cen B & 19575 &  HD 139211 & 1520 &   HD 207129 & 494 &                HD 176986 & 324 \\
* tau Cet & 12132 &    HD  85512 & 1300 &   HD   1461 & 487 &                BD+05  1668 & 320 \\
* bet Pic & 9098 &     *  61 Vir & 1279 &   HD  40307 & 478 &                *  51 Peg & 309 \\
* e Eri & 7269 &       HD  49933 & 1234 &   Proxima Cen & 457 &    Ross  128 & 307 \\
* del Pav & 6007 &     * del Eri & 1217 &   HD 144628 & 452 &                * z Cen & 305 \\
* alf Cen A & 5838 &   HD 199288 & 1199 &   * rho Pup & 447 &                HD  95456 & 303 \\
* alf CMi & 5710 &     HD  46375 & 1165 &   * zet02 Ret & 417 &              * lam02 For & 300 \\
*  18 Sco & 5510 &     V* V816 Cen & 1123 & HD 131977 & 416 &                G 268-38 & 296 \\
* eta Boo & 5175 &     HD  65907 & 1050 &   HD 133112 & 412 &                * eta02 Hyi & 294 \\
* alf Cir & 4970 &     * eps Pav & 887 &    * d Sco & 403 &                  HD 161098 & 290 \\
* eps Ind & 4717 &     HD 109200 & 864 &    CD-45  5378 & 402 &              HD 134606 & 290 \\
* zet Tuc & 3927 &     HD 217987 & 845 &    HD   4308 & 391 &                * d Boo & 288 \\
NO NAME & 3077 &HD 154577 & 840 &    HD  44195 & 387 &                HD  36003 & 285 \\
* mu. Ara & 3018 &     * omi02 Eri & 781 &  HD  33793 & 386 &                HD 154088 & 280 \\
* bet Hyi & 2886 &     * eps Eri & 768 &    HD 109536 & 386 &                BD+01   316 & 279 \\
* gam Pav & 2625 &     HD  69830 & 755 &    HD 215152 & 375 &                HD 210918 & 278 \\
* bet Aql & 2398 &     HD 114613 & 747 &    * tet Lup & 370 &                HD  39194 & 277 \\
* iot Hor & 2184 &     HD 189567 & 659 &    BD+10  1799 & 369 &              *  30 Mon & 276 \\
* tau PsA & 2183 &     * nu.02 Lup & 657 &  HD  96700 & 366 &                HD  38858 & 274 \\
HD 192310 & 1835 &     * pi. Men & 557 &    HD  45184 & 355 &                HD 156384C & 273 \\
*  70 Oph & 1797 &     HD 102365 & 554 &    HD  93396 & 342 &                BD-15  6290 & 270 \\
*  94 Cet & 1766 &     HD 172555 & 536 &    HD 134060 & 339 &                HD  31527 & 268 \\
*  32 Aqr & 1762 &     CD-40  9712 & 536 &  HD  10180 & 336 &                HD 223171 & 267 \\
* 171 Pup & 1693 &     HD  72673 & 531 &    Barnard star & 335 &      HD 215456 & 267 \\
* alf PsA & 1593 &     HD  59468 & 531 &    V* YZ Cet & 334 &                HD 106315 & 265 \\
\hline
\end{tabular}
\caption{Most 100 observed object by HARPS}
\label{t_obs}
\end{table}
\clearpage

\begin{table}[ht!]
\scriptsize
\begin{tabular}{|l|l|}
\hline
TABLE COLUMN & DESCRIPTION AND UNITS\\
\hline
\hline
adp\_dpid & ESO PHASE3 ADP file name \\
raw\_dpid & ESO RAW file name \\
\hline
progid & ESO program identification \\
dpr\_tech & observation technique (Echelle or Echelle with Iodine Absorption Cell) \\
dpr\_type & three fields: object type in fiber A, object in fiber B, and spectral type or lamp in fiber A \\
ins\_mode & High RV Accuracy Mode (HAM), High Efficiency Mode (EGGS) \\
\hline
tel\_object & Target designation as given by the astronomer \\
tel\_targ\_alpha & Target RA coordinate as given by the astronomer (deg) \\
tel\_targ\_delta & Target DEC coordinate as given by the astronomer (deg) \\
date\_obs & Date and time of beginning of observation \\
mjd\_obs & MJD start of observation (d) \\
exptime & Total integration time (s) \\
ins\_det1\_tmmean & Normalized mean exposure time on fiber A (unit-less) \\
\hline
drs\_bjd & Barycentric Julian Day (d) \\
drs\_version\_number & HARPS DRS version number \\
drs\_snr & Median signal to noise ratio of the spectra \\
drs\_berv & Barycentric Earth Radial Velocity (BERV) (km/s) \\
drs\_mask & Name of the spectral template for CCF measurement (G2,K0,K5,M2,M4) \\
drs\_ccf\_rvc & Barycentric RV (drift corrected) (km/s) \\
drs\_ccf\_rv & Barycentric RV (no drift correction) (km/s) \\
drs\_bis\_rv & Bisector mean velocity (km/s) \\
drs\_dvrms & Estimated RV uncertainty (m/s) \\
drs\_ccf\_contrast & Contrast of CCF (\%) \\
drs\_ccf\_fwhm & FWHM of CCF (km/s) \\
drs\_ccf\_noise & Photon noise on CCF RV (km/s) \\
drs\_bis\_span & Bisector velocity span (km/s) \\
drs\_drift\_noise & Thorium lamp Drift photon noise (m/s) \\
drs\_drift\_rv\_used & Used RV Drift (m/s) \\
\hline
H\_I\_6562\_rv & Hydrogen I 6562 AA line Radial Velocity (km/s) \\
H\_I\_6562\_rv\_error & Hydrogen I 6562 AA line Radial Velocity Error (km/s) \\
H\_I\_6562\_fit\_chi2 & Hydrogen I 6562 AA line Gaussian FIT CHI2 \\
\hline
main\_id\_simbad & SIMBAD main ID \\
sp\_type\_simbad & SIMBAD spectral type \\
main\_type\_simbad & SIMBAD main type \\
other\_types\_simbad & SIMBAD other types \\
ra\_simbad & SIMBAD RA (deg) \\
dec\_simbad & SIMBAD DEC (deg) \\
radvel\_simbad & SIMBAD radial velocity (km/s) \\
radvel\_err\_simbad & SIMBAD radial velocity error (km/s) \\
plx\_simbad & SIMBAD parallax (mas) \\
plx\_err\_simbad & SIMBAD parallax error (mas) \\
pmra\_simbad & SIMBAD proper motion in RA (mas/yr) \\
pmdec\_simbad & SIMBAD proper motion in DEC (mas/yr) \\
pmra\_err\_simbad & SIMBAD proper motion error in RA (mas/yr) \\
pmdec\_err\_simbad & SIMBAD proper motion error in DEC (mas/yr) \\
pm\_err\_pa\_simbad & SIMBAD proper motion error position angle (deg) \\
Bmag\_simbad & SIMBAD B magnitude (mag) \\
Vmag\_simbad & SIMBAD V magnitude (mag) \\
Jmag\_simbad & SIMBAD J magnitude (mag) \\
Hmag\_simbad & SIMBAD H magnitude (mag) \\
Kmag\_simbad & SIMBAD K magnitude (mag) \\
angDist\_simbad & angular distance between the observed target and the SIMBAD source (arcsec) \\
\hline
\end{tabular}
\caption{Description of the columns in the catalog.
The data from SIMBAD corresponds to the version accessible online as of June 1, 2023.}
\label{tab01}
\end{table}

\clearpage

\section{Data Quality}

\begin{figure}[ht!]
 \centering
 \includegraphics[width=\columnwidth]{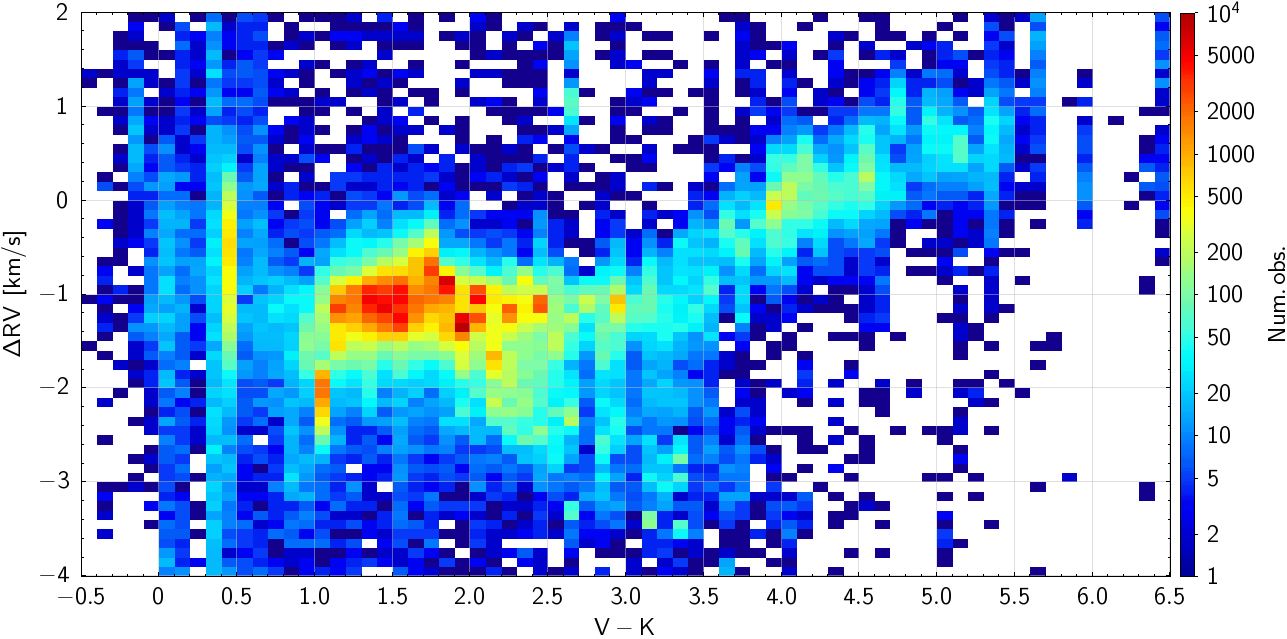}
 \caption{Difference in RV between HARPS DRS and H$\alpha$, as function of color V-K. The color bar represents the density of observations in each bin of the 2D histogram.}
 \label{fig:07}

 \vspace{10pt}

 \includegraphics[width=\columnwidth]{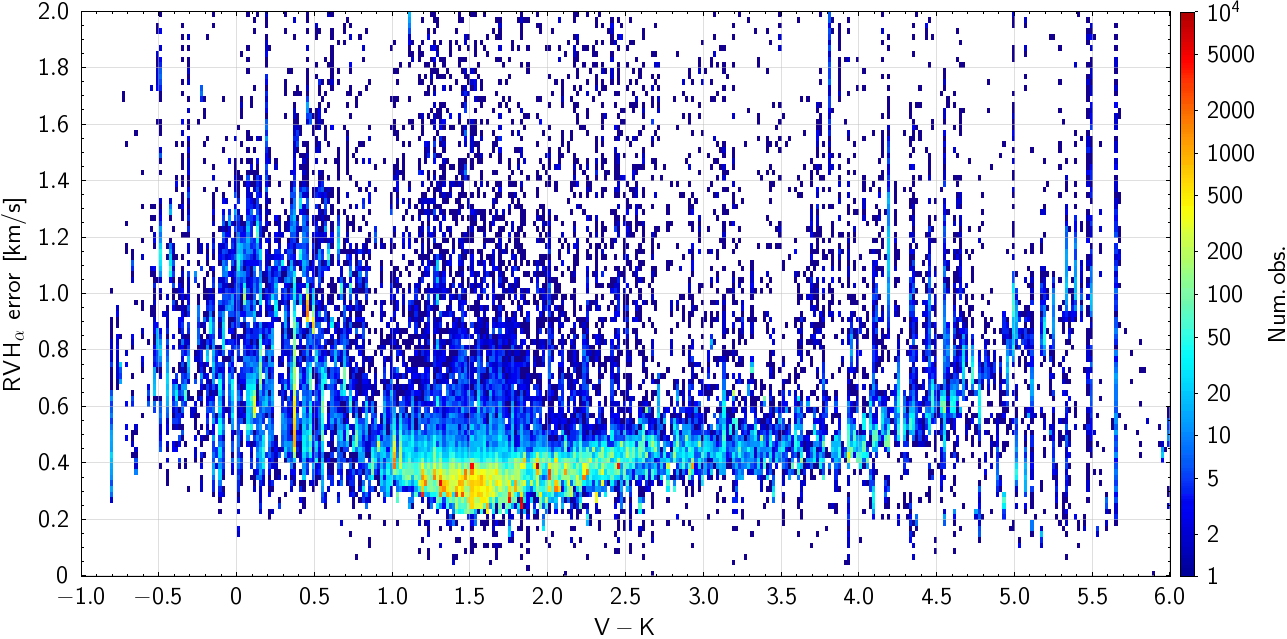}
 \caption{Error in RV H$\alpha$, as function of color V-K. The color bar represents the density of observations in each bin of the 2D histogram.}
 \label{fig:08}
\end{figure}

The measurements of RV from HARPS DRS and H$\alpha$ line generally show good agreement, within 2 km/s, however the difference between the two RV values is a function of stellar color, ranging from -2 km/s for F stars to +1 km/s for M stars (see Fig. \ref{fig:07}).

This difference is largely attributed to the intrinsic properties of the H$\alpha$ line generation under non-LTE conditions and the presence of convective motions in stellar atmospheres.

\begin{figure}[p]
 \centering
 \begin{adjustbox}{center}
   \rotatebox{-90}{\includegraphics[width=20cm]{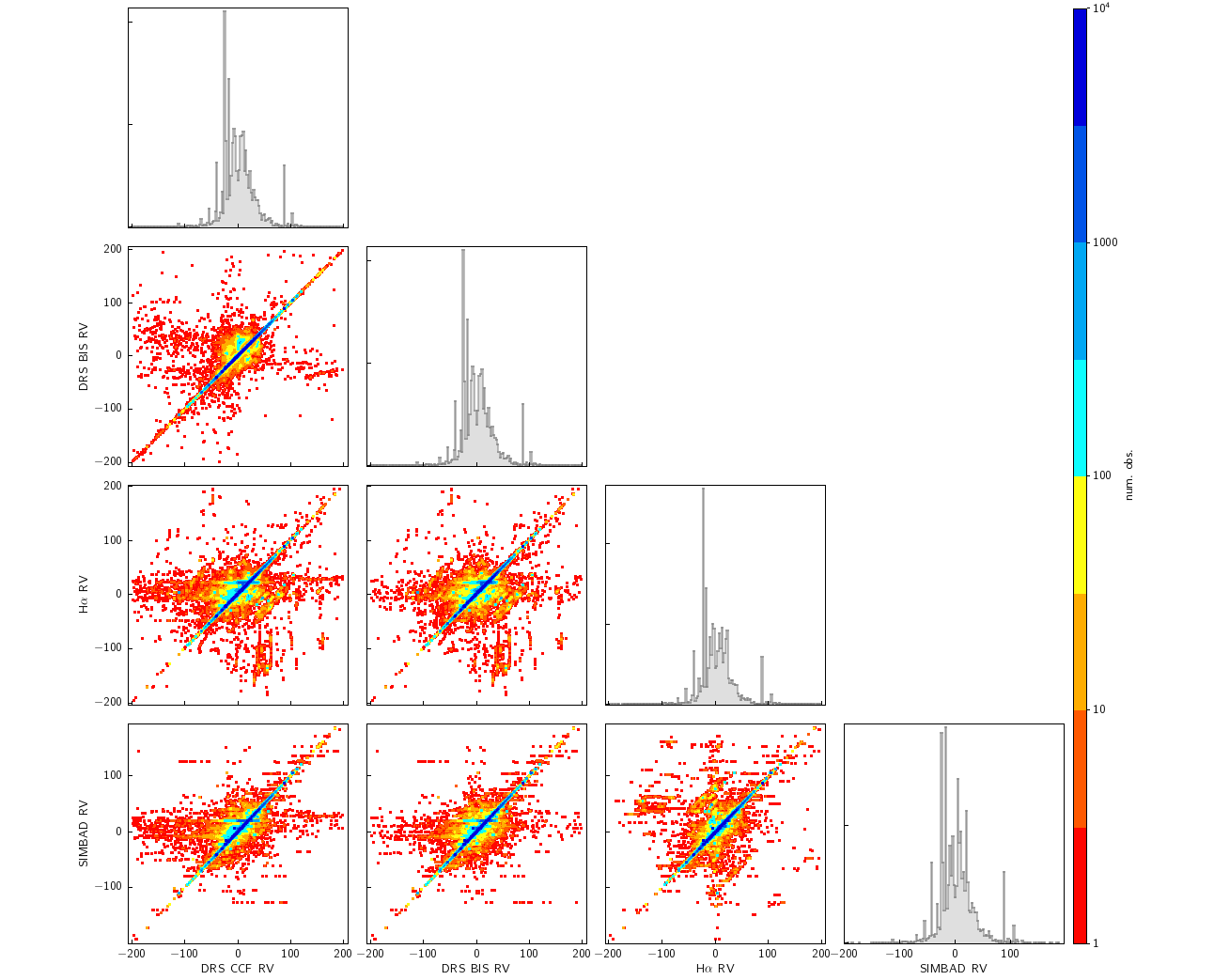}}
 \end{adjustbox}
 \caption{Comparison of the various RVs collected in this catalog. The outliers from the diagonal are typically spectra with low SNR.
 The units of measure in the axis is km/s.
 }
 \label{fig:08a}
\end{figure}

Due to their high quality and small errors, RVs determined with HARPS DRS can be used for exoplanet searches and asteroseismology. In contrast, H$\alpha$ RVs have typical errors of 300 m/s and are not suitable for exoplanet searches.
The distribution of radial velocity uncertainties in H$\alpha$ with respect to the color V-K is presented in Figure \ref{fig:08}. It is observed that the majority of the measurements have errors less than 500 m/s. Notably, the errors increase for stars with redder color indices (higher V-K values). This trend can be attributed to the diminishing intensity of the H$\alpha$ as the stellar temperature decrease (i.e. as V-K increase). Such a decrease in line intensity likely contributes to the higher RV measurement uncertainties for these stars.
A comparison of all the RVs (CCF, bisectors, H$\alpha$, SIMBAD)
compiled in this catalog, is presented in Fig.~\ref{fig:08a}.

\subsection{Known issues}

In some cases, the HARPS DRS values for the radial velocities are not reliable. We could identify the following reasons for this behavior that the catalog user need to take in to account:
\begin{itemize}
\item Discrepant RV from those available in the literature: in this case probably the HARPS DRS failed to converge because either the observer provided an incorrect input RV, or an incorrect mask was used, or the spectra was not of good quality (low SNR, blaze functions not completely removed).

\item RV value too large: since almost all the observations of HARPS are stars within our Galaxy, radial velocities of stars under normal circumstances cannot exceed the 1\,000 km/s in absolute value. In these cases is most probable that the HARPS DRS fit did not converge.

\item Radial velocities for stars hotter than F8 were reduced by the HARPS DRS with the G2 mask. This fact especially for early type stars (O,B,A) may induce non convergence in the HARPS DRS results, and hence reduced quality of the radial velocity.

\item Radial velocities of binary stars with small angular separation (less than 2 arcsec) under poor seeing condition or at large airmass may suffer from contamination from the other component, making RV determination uncertain.

\item During the period 2003--2004, HARPS operated also with an absorption Iodine cell.
These observations can be identified in the catalog via the column dpr\_type and with the value "ECHELLE,ABSORPTION-CELL", and amounts to 1\,487 in total.
In the Iodine self-calibration method the DRS pipeline does spectrum extraction and applies the wavelength calibration. Considering that I2
lines "pollute" the spectra, the DRS radial velocity should be considered as an approximate value. The users that want to use these spectra need to be aware that the I2 absorption lines were not removed from the stellar spectra.

\end{itemize}

The RV measured with the H$\alpha$ line sometimes provides values that are too large and this is due to the fact that the fit does not converge. These particular cases could be identified by largest values of the reduced chi squared.

\clearpage

\section{Acknowledgments}

This research has made use of the SIMBAD database, operated at CDS, Strasbourg, France.

\section{Data policy}

\noindent
Any publication making use of this data, whether obtained from the ESO archive or via third parties, must include the following acknowledgment:
~\\
~\\
\noindent
\textit{Based on data obtained from the ESO Science Archive Facility with DOI(s):}\\
\url{https://doi.org/10.18727/archive/33}
~\\

\noindent
Science data products from the ESO archive may be distributed by third parties, and disseminated via other services, according to the terms of the Creative Commons Attribution 4.0 International license. Credit to the ESO origin of the data must be acknowledged, and the file headers preserved.
~\\

\clearpage
\appendix
\section*{APPENDIX}

\section{Problems found in the metadata of object names and coordinates}
\label{app_metadata}

When preparing an observation, astronomers must complete the Observing Block (OB) with essential details of the target, including its name, equatorial coordinates, proper motion, parallax, and radial velocity. These parameters are recorded in both RAW and ADP files.
Here, we present the keyword values in the FITS files that represent the astronomer's input for an observation of the star Procyon ($\alpha$ CMi).

{\small
\begin{verbatim}
HIERARCH ESO OBS TARG NAME = 'Procyon' / OB target name
HIERARCH ESO TEL TARG ALPHA = 73918.118 / Alpha coordinate for the target
HIERARCH ESO TEL TARG COORDTYPE = 'M' / Coordinate type (M=mean A=apparent)
HIERARCH ESO TEL TARG DELTA = 51329.980 / Delta coordinate for the target
HIERARCH ESO TEL TARG EPOCH = 2000.000 / Epoch
HIERARCH ESO TEL TARG EPOCHSYSTEM = 'J' / Epoch system (default J=Julian)
HIERARCH ESO TEL TARG EQUINOX = 2000.000 / Equinox
HIERARCH ESO TEL TARG PARALLAX = 0.000 / Parallax
HIERARCH ESO TEL TARG PMA = -0.720000 / Proper Motion Alpha
HIERARCH ESO TEL TARG PMD = -1.030000 / Proper motion Delta
HIERARCH ESO TEL TARG RADVEL = -3.200 / Radial velocity
\end{verbatim}}

\noindent
{\tt TARG ALPHA} and {\tt TARG DELTA} are presented in pseudo-sexagesimal format,
and their values must be read as follows:\\
{\tt TARG ALPHA}=73918.118 is RA=07:39:18.118\\
{\tt TARG DELTA}=51329.980 is DEC=+05:13:29.980.

~\\
\noindent
For reference we report the characteristics of Procyon retrieved from SIMBAD.

\begin{table}[h!]
\begin{tabular}{|l|l|}
\hline
Name & Procyon, $\alpha$ CMi \\
RA & 07:39:18.11950 \\
DEC & +05:13:29.9552 \\
$\mu_\alpha$ & -0.71459 \\
$\mu_\delta$ & -1.03680 \\
RV & -4.505280 \\
Spectral type & F5IV-V+DQZ \\
Object type & Spectroscopic Binary \\
V mag & 0.37 \\
\hline
\end{tabular}
\end{table}

\subsection{Object Names}

The nomenclature for astronomical targets can vary significantly among astronomers, often leading to inconsistencies. One common issue is the use of different astronomical catalog names for the same target. Additionally, typographical errors and alternative names further complicate the matter. Here we present some illustrative examples to highlight these problematics.

\begin{itemize}
 \item alternative names:\\
 {\tt HD 128621}, {\tt HR 5460}, {\tt ALPHACENB},
 {\tt ALPCENB}, all of them used as OBJECT name for $\alpha$ Cen B
 \item typographical errors :\\
 {\tt GJ-550} instead of {\tt GJ 550}\\
 {\tt BD--4-3208} instead of {\tt BD-04 3208}\\
 {\tt HD-209625} instead of {\tt HD 209625}
 \item ambiguous names:\\
 {\tt V V Cen} = is {\tt V* VV Cen} or {\tt V* V Cen} ?\\
 {\tt TYCHO} instead of {\tt Moon crater Tycho}
 \item misspellings: \\
 {\tt HD 550} instead of {\tt GJ 550} (confused the catalog)\\
 {\tt 493.01} instead of {\tt TOI 493}
 \item invented nicknames: \\
 {\tt TCHOUBIDOO1} instead of {\tt CD-27 10695}\\
 {\tt U1} instead of {\tt OGLE LMC-CEP-79}\\
 {\tt 620} instead of {\tt 2MASS J16591164-5244429}
\end{itemize}

\subsection{Equatorial Coordinates}

The RA and DEC keywords in the FITS files represent the pointing coordinates calculated by the Telescope Control Software (TCS) at the time of observation, rather than the user-provided input coordinates during OB preparation. These calculations incorporate various factors affecting the observation, including time, date, wavelength, humidity, pressure, telescope's geographic location (longitude, latitude, altitude), the telescope's mechanical pointing model, differential velocities for Solar System objects, and the target's specific details (coordinate system, epoch, equinox, parallax, radial velocity, and proper motions).

The RA and DEC coordinates are calculated in the FK5 J2000 system, and includes  all necessary corrections such as precession, nutation, polar motion, target proper motion, atmospheric refraction, and adjustments for the telescope's mechanical pointing model. The recorded coordinates of the same target exhibit variability due to several factors:
\begin{itemize}
\item adjustments for Earth motion (precession, nutation, polar motion),
\item the target proper motion
\item the specific mechanical adjustments needed for each observation for compensating telescope flexures, which vary depending on the telescope altazimuthal coordinates
\item the evolving mechanical condition (aging) of the telescope over time
\end{itemize}
This means that even for repeated observations of the same target,
the recorded coordinates can differ due to these variable factors.
Furthermore, the extent of this coordinate scatter differs from target to target and changes over time, influenced by factors such as the aging of the telescope's mechanics, maintenance interventions, or natural events like earthquakes.

When preparing the Observing Blocks (OB), the coordinates input by the observer may be prone to various errors. Our study of the HARPS metadata has revealed several recurrent cases, which we detail in the following list.

\begin{itemize}
\item ICRS 1991.5 coordinates specified as FK5 and epoch 2000. In this case it means that the star proper motions over roughly 8.5 years aren't accounted for.
\item B1950 coordinates specified as FK5 and epoch 2000. In this case the star proper motions and Earth precession have shifted the star position of many arcmin.
\item ICRS or FK5 coordinates with proper motions calculated for the epoch of observation, while epoch of the coordinates inserted as J2000.
\item Proper motions not inserted.
\item Proper motions specified in mas/yr instead of arcsec/yr,
typically happening for Hipparcos stars.
\end{itemize}

\section{On the fiber size and light scrambling}
\label{app_fiber}

In this section, we aim to clarify some regarding the optical fiber replacement of HARPS in 2015.
For more detailed analysis and technical insights, we refer the readers to the following articles: \cite{2013A&A...549A..49B}, \cite{2018MNRAS.475.3065G}.

\subsection{Importance of light scrambling}
    \begin{itemize}
        \item In spectroscopy, detecting minute shifts in spectral lines is crucial for measuring radial velocities of stars. These measurements can be extremely sensitive to how light enters and travels through the fiber.
        \item If the light entering the fiber retains any patterns or uneven distributions from its source, these irregularities can mimic or mask true spectral shifts, leading to inaccurate measurements.
        \item Effective light scrambling mixes the light paths within the fiber, smoothing out these irregularities, and ensuring that the light exiting the fiber is evenly distributed, regardless of its initial distribution. This uniformity is critical for accurate spectroscopic measurements.
    \end{itemize}

\subsection{
Why octagonal fibers redistribute light more evenly than circular fibers
}
    \begin{itemize}
        \item The shape of the fiber affects how light is reflected internally. In a circular fiber, reflections are more predictable and can reinforce uneven distributions of light.
        \item In contrast, the flat sides and angles of an octagonal fiber create a more complex pattern of reflections. Each time light hits one of these flat sides at an angle, it's reflected in a different direction compared to a circular fiber.
        \item This complexity in reflections causes the light paths to intersect and overlap more frequently, leading to a more uniform distribution by the time the light exits the fiber.
    \end{itemize}

\subsection{
Why positional and intensity fluctuations are reduced using octagonal fibers vs. circular fibers
}
    \begin{itemize}
        \item In a circular fiber, slight misalignments or changes in the intensity of the light source can lead to significant variations in how the light travels through the fiber, affecting the output signal.
        \item The octagonal shape, with its more complex internal reflection pattern, acts to average out these variations. This means that small changes in the position or brightness of the light source result in less variation in the light distribution at the output.
        \item Essentially, the octagonal fiber's geometry acts as a buffer against minor inconsistencies in light input, leading to a more stable and consistent output, which is crucial for precision measurements in spectroscopy.
    \end{itemize}

\section{Software for retrieving ADP, ANCILLARY and RAW file names}
\label{app_code}

For creating a list of correspondences between ADP science files, ancillary TAR files (containing CCF and BIS measurements), and RAW files we build a script in python that perform TAP\footnote{\url{https://www.ivoa.net/documents/TAP}} queries from ESO TAP server\footnote{\url{http://archive.eso.org/tap_cat}}. It queries different tables (dbo.raw, ivoa.obscore, phase3v2.files) and processes them to create a unified table. The script includes steps for querying data, filtering based on release dates and filtering based on latest available version. Finally it merge the different data sources into a final table. The script utilizes these libraries: Astropy, PyVO, Pandas, and Numpy.

\begin{lstlisting}
#!/usr/bin/env python3
import sys
import csv
from datetime import datetime
import time
from astropy.table import Table
from astropy.time import Time
import pyvo as vo
import pandas as pd
import numpy as np

# Function to convert ISO date string to MJD
def convert_to_mjd(date_str):
    return Time(date_str).mjd

print()
print("HARPS ESO TAP QUERY FOR RETRIEVING ADP, AUX, RAW FILE NAMES")
print()

# Get today's date
today = datetime.now().date()

# Define the TAP service URL
tap_service_url = "http://archive.eso.org/tap_obs"

# Create a TAP service object
tap_service = vo.dal.TAPService(tap_service_url)

query_raw = """
select *
from dbo.raw
where instrument = 'HARPS' and
dp_tech like 'ECHELLE%' and
dp_type like 'STAR%' and
prog_id != '1102.D-0954(A)' and
exposure>0
"""

query_adp = """
select *
FROM ivoa.obscore
WHERE instrument_name = 'HARPS' AND
obs_collection = 'HARPS' AND
dataproduct_type = 'spectrum' AND
proposal_id != '1102.D-0954(A)' AND
snr > 0 AND
t_exptime > 0
ORDER BY t_min ASC
"""

query_sci = """
select *, eso_substring(name,1,29) as dpid1,
          eso_substring(archive_id,5,31) as dateadp
from phase3v2.files
where name like 'HARPS%' and
category = 'SCIENCE.SPECTRUM'
order by dpid1
"""

query_aux = """
select *, eso_substring(name,1,29) as dpid1,
        eso_substring(archive_id,5,31) as dateadp
from phase3v2.files
where name like 'HARPS%' and
category = 'ANCILLARY.HARPSTAR'
order by dpid1
"""

start_time = time.time()  # start time
print('Executing TAP query on raw files')
print(query_raw)
tap_raw = tap_service.search(query_raw, maxrec=100000000)
elapsed_time = time.time() - start_time  # Calculate elapsed time
print(f'N raw {len(tap_raw)}')
print(f'Exec time: {elapsed_time:.2f} s')  # elapsed time
print()
print()

start_time = time.time()  # start time
print('Executing TAP query on adp files')
print(query_adp)
tap_adp = tap_service.search(query_adp, maxrec=100000000)
elapsed_time = time.time() - start_time  # Calculate elapsed time
print(f'N adp {len(tap_adp)}')
print(f'Exec time: {elapsed_time:.2f} s')  # elapsed time
print()
print()

start_time = time.time()  # start time
start_time0 = start_time
print('Executing TAP query on sci files')
print(query_sci)
tap_sci = tap_service.search(query_sci, maxrec=100000000)
elapsed_time = time.time() - start_time  # Calculate elapsed time
print(f'N sci {len(tap_sci)}')
print(f'Exec time: {elapsed_time:.2f} s')  # elapsed time
print()
print()

start_time = time.time()  # start time
print('Executing TAP query on aux files')
print(query_aux)
tap_aux = tap_service.search(query_aux, maxrec=100000000)
elapsed_time = time.time() - start_time  # Calculate elapsed time
print(f'N aux {len(tap_aux)}')
print(f'Exec time: {elapsed_time:.2f} s')  # elapsed time
print()
print()

start_time = time.time()  # start time
print('Convert the TAPResults to DataFrames for raw and adp')
df_raw = tap_raw.to_table().to_pandas()
df_adp = tap_adp.to_table().to_pandas()
elapsed_time = time.time() - start_time  # Calculate elapsed time
print(f'Exec time: {elapsed_time:.2f} s')  # elapsed time

start_time = time.time()  # start time
print('Convert release_date and obs_release_date columns to datetime')
df_raw['release_date'] = pd.to_datetime(df_raw['release_date'])
df_adp['obs_release_date'] = pd.to_datetime(df_adp['obs_release_date'])
elapsed_time = time.time() - start_time  # Calculate elapsed time
print(f'Exec time: {elapsed_time:.2f} s')  # elapsed time

start_time = time.time()  # start time
print('Filter the DataFrames raw and adp')
table_raw = df_raw[df_raw['release_date'].dt.date <= today]
table_adp = df_adp[df_adp['obs_release_date'].dt.date <= today]
elapsed_time = time.time() - start_time  # Calculate elapsed time
print(f'N raw after removing files availables in the future {table_raw.shape[0]}')
print(f'N adp after removing files availables in the future {table_adp.shape[0]}')
print(f'Exec time: {elapsed_time:.2f} s')  # elapsed time

start_time = time.time()  # start time
print('Create the DataFrames for sci and aux')
table_sci = tap_sci.to_table().to_pandas()
table_aux = tap_aux.to_table().to_pandas()
elapsed_time = time.time() - start_time  # Calculate elapsed time
print(f'Exec time: {elapsed_time:.2f} s')  # elapsed time

start_time = time.time()  # start time
print('Add dateadp column to table_sci and table_aux')
# Extract relevant substring and convert to MJD for table_sci
sci_date_strings = table_sci['archive_id'].str[4:27]
table_sci['dateadp'] = sci_date_strings.apply(convert_to_mjd)
# Extract relevant substring and convert to MJD for table_aux
aux_date_strings = table_aux['archive_id'].str[4:27]
table_aux['dateadp'] = aux_date_strings.apply(convert_to_mjd)

elapsed_time = time.time() - start_time  # Calculate elapsed time
print(f'Exec time: {elapsed_time:.2f} s')  # elapsed time


start_time = time.time()  # start time
print('Sort by dateadp in descending order')
table_sci = table_sci.sort_values(by='dateadp', ascending=False)
table_aux = table_aux.sort_values(by='dateadp', ascending=False)
elapsed_time = time.time() - start_time  # Calculate elapsed time
print(f'Exec time: {elapsed_time:.2f} s')  # elapsed time

start_time = time.time()  # start time
print('Keep only the most recent record for each dpid1')
table_sci = table_sci.drop_duplicates(subset='dpid1', keep='first')
table_aux = table_aux.drop_duplicates(subset='dpid1', keep='first')
elapsed_time = time.time() - start_time  # Calculate elapsed time
print(f'Exec time: {elapsed_time:.2f} s')  # elapsed time

start_time = time.time()  # start time
print('Rename columns in tables by appending extensions')
table_raw.columns = [f'{col}_raw' for col in table_raw.columns]
table_adp.columns = [f'{col}_adp' for col in table_adp.columns]
table_sci.columns = [f'{col}_sci' for col in table_sci.columns]
table_aux.columns = [f'{col}_aux' for col in table_aux.columns]
elapsed_time = time.time() - start_time  # Calculate elapsed time
print(f'Exec time: {elapsed_time:.2f} s')  # elapsed time

start_time = time.time()  # start time
print('Match table_raw and table_sci')
table_temp1 = pd.merge(table_raw, table_sci, left_on='dp_id_raw', right_on='dpid1_sci')
elapsed_time = time.time() - start_time  # Calculate elapsed time
print(f'Exec time: {elapsed_time:.2f} s')  # elapsed time

start_time = time.time()  # start time
print('Match table_temp1 and table_aux')
table_temp2 = pd.merge(table_temp1, table_aux, left_on='dp_id_raw', right_on='dpid1_aux')
elapsed_time = time.time() - start_time  # Calculate elapsed time
print(f'Exec time: {elapsed_time:.2f} s')  # elapsed time

start_time = time.time()  # start time
print('Match table_temp2 and table_adp')
table_final = pd.merge(table_temp2, table_adp, left_on='archive_id_sci', right_on='dp_id_adp')
elapsed_time = time.time() - start_time  # Calculate elapsed time
print(f'Exec time: {elapsed_time:.2f} s')  # elapsed time

num_rows = table_final.shape[0]
print('Saving results to harps_raw_adp_aux.table.csv')
print(f"Number of rows in harps_raw_adp_aux.table.csv: {num_rows}")
table_final.to_csv('harps_raw_adp_aux.table.csv', index=False)
elapsed_time = time.time() - start_time0  # Calculate elapsed time
print(f'Total exec time: {elapsed_time:.2f} s')  # elapsed time
print()
\end{lstlisting}


\section{Solar System bright object names}
\label{app_ss}

To identify HARPS observations associated with bright Solar System objects, 
we compiled a list of 220 names (Table \ref{tab_ss}), including the Sun, planetary satellites, 
and the first 100 numbered asteroids. The list features the names of these o
bjects in English, along with their variants in the languages spoken in ESO 
member countries. Observations with the keyword OBJECT matching these 
names were excluded from the list of stellar observations.

\begin{table}[h!]
\centering
\scriptsize
\begin{tabular}{|l|l|l|l|l|}
\hline
Adrastea & Egeria & Julia & Mond & Semele \\
Aegina & Elara & Juno & Moon & Sinope \\
Aegle & Elpis & Jupiter & Naiad & Slonce \\
Aglaja & Enceladus & Kalliope & Nemausa & Sol \\
Alexandra & Epimetheus & Kallisto & Neptun & Sole \\
Alkmene & Erato & Kalypso & Neptune & Soleil \\
Amalthea & Eugenia & Kerberos & Neptuno & Sonne \\
Amphitrite & Eunomia & Klio & Neptunus & Styx \\
Ananke & Euphrosyne & Klotho & Nereid & Sun \\
Angelina & Europa & Klytia & Nettuno & Sylvia \\
Antiope & Eurydike & Laetitia & Netuno & Telesto \\
Arethusa & Eurynome & Land & Niobe & Terpsichore \\
Ariadne & Euterpe & Landa & Nix & Terra \\
Ariel & Feronia & Larissa & Nysa & Terrain \\
Asia & Fides & Leda & Oberon & Tethys \\
Astraea & Flora & Leto & Ophelia & Thalassa \\
Atalante & Fortuna & Leukothea & Pales & Thalia \\
Atlas & Freia & Lsiezyc & Pallas & Thebe \\
Aurora & Frigga & Lua & Pan & Themis \\
Ausonia & Galatea & Luna & Pandora & Thetis \\
Beatrix & Ganimede & lune & Panopaea & Thisbe \\
Bellona & Ganimedes & Lutetia & Parthenope & Tierra \\
Bianca & Ganymed & Lysithea & Pasiphae & Tita \\
Calisto & Ganymede & Maan & Phobos & Titan \\
Callisto & Ganymedes & Maja & Phocaea & Titania \\
Calypso & Giove & Mane & Phoebe & Titano \\
Carme & Grunt & Mars & Plutao & Tritao \\
Caronte & Harmonia & Marte & Pluto & Triton \\
Ceres & Hebe & Massalia & Pluton & Tritone \\
Charon & Hekate & Melete & Plutone & Tryton \\
Circe & Helene & Melpomene & Polydeuces & Tytan \\
Concordia & Hesperia & Mercure & Polyhymnia & Umbriel \\
Cordelia & Hestia & Mercurio & Pomona & Undina \\
Cybele & Himalia & Mercurius & Prometheus & Uran \\
Danae & Hippocamp & Mercury & Proserpina & Urania \\
Daphne & Hygiea & Merkur & Proteus & Urano \\
Deimos & Hyperion & Merkurius & Psyche & Uranus \\
Despina & Ianthe & Merkury & Puck & Venere \\
Diana & Iapetus & Methone & Rhea & Venus \\
Dike & Irene & Metis & Sappho & Vesta \\
Dione & Iris & Mimas & Saturn & Victoria \\
Doris & Isis & Minerva & Saturne & Virginia \\
Earth & Janus & Miranda & Saturno & Wenus \\
Echo & Jowisz & Mnemosyne & Saturnus & Zon \\
\hline
\end{tabular}
\caption{Names of Solar System bodies used for filtering stellar observations.}
\label{tab_ss}
\end{table}

\end{document}